\documentclass[Crown,sagev,times,doublespace]{sagej}

\usepackage{moreverb,url}
\usepackage{afterpage}
\usepackage[colorlinks,bookmarksopen,bookmarksnumbered,citecolor=red,urlcolor=red]{hyperref}
\usepackage{rotating}
\usepackage{color,colortbl}
\usepackage{changepage}
\definecolor{Gray}{gray}{0.9}

\newcommand\BibTeX{{\rmfamily B\kern-.05em \textsc{i\kern-.025em b}\kern-.08em
T\kern-.1667em\lower.7ex\hbox{E}\kern-.125emX}}

\newcommand{\bcx}{{\bf X}}
\newcommand{\bcy}{{\bf Y}}

\newcommand{\bcv}{{\bf V}}
\newcommand{\bcw}{{\bf W}}
\newcommand{\bci}{{\bf I}}

\newcommand{\bcr}{{\bf R}}
\newcommand{\bcm}{{\bf M}}

\newcommand{\bw}{{\bf w}}

\newcommand{\bmu}{\mbox{\boldmath $\mu$}}

\newcommand{\bfgamma}{\mbox{\boldmath $\gamma$}}

\newcommand{\bfeta}{\mbox{\boldmath $\eta$}}

\newcommand{\bdelta}{\mbox{\boldmath $\Delta$}}

\newcommand{\balpha}{\mbox{\boldmath $\alpha$}}

\newcommand{\bsigma}{{\bf \Sigma}}

\newcommand{\bomega}{\mbox{\boldmath $\Omega$}}
\newcommand{\bfomega}{\mbox{\boldmath $\omega$}}

\theoremstyle{plain}

\theoremstyle{definition}
\usepackage[]{graphicx}
\chardef\bslash=`\\ % p. 424, TeXbook

\makeatletter
\newcommand*{\rom}[1]{\expandafter\@slowromancap\romannumeral #1@}
\makeatother

\def\volumeyear{2023}

\begin{document}

\runninghead{Blette, Halpern, Li, and Harhay}

\title{Assessing treatment effect heterogeneity in the presence of missing effect modifier data in cluster-randomized trials}

\author{Bryan S. Blette\affilnum{1}, Scott D. Halpern\affilnum{2,3}, Fan Li\affilnum{4,5,*}, and Michael O. Harhay\affilnum{2,3,*}}

\affiliation{\affilnum{1}Department of Biostatistics, Vanderbilt University Medical Center, Nashville, TN, USA\\
\affilnum{2}Department of Biostatistics, Epidemiology, and Informatics, Perelman School of Medicine, University of Pennsylvania, Philadelphia, PA, USA\\
\affilnum{3}Clinical Trials Methods and Outcomes Lab, PAIR (Palliative and Advanced Illness Research) Center, Perelman School of Medicine, University of Pennsylvania, Philadelphia, PA, USA\\
\affilnum{4}Department of Biostatistics, Yale School of Public Health, New Haven, CT, USA\\
\affilnum{5}Center for Methods in Implementation and Prevention Science, Yale University, New Haven, CT, USA}

\corrauth{Bryan Blette, 2525 West End Ave, 11118A, Nashville, TN 37203}

\email{bryan.blette@vumc.org\\ 
*F. Li and M.O. Harhay are co-senior authors}

\begin{abstract}
Understanding whether and how treatment effects vary across subgroups is crucial to inform clinical practice and recommendations. Accordingly, the assessment of heterogeneous treatment effects (HTE) based on pre-specified potential effect modifiers has become a common goal in modern randomized trials. However, when one or more potential effect modifiers are missing, complete-case analysis may lead to bias and under-coverage. While statistical methods for handling missing data have been proposed and compared for individually randomized trials with missing effect modifier data, few guidelines exist for the cluster-randomized setting, where intracluster correlations in the effect modifiers, outcomes, or even missingness mechanisms may introduce further threats to accurate assessment of HTE. In this article, the performance of several missing data methods 
%(complete case analysis, single imputation, multiple imputation, multilevel multiple imputation [MMI], and Bayesian MMI) 
are compared through a simulation study of cluster-randomized trials with continuous outcome and missing binary effect modifier data, and further illustrated using real data from the 
%Thereafter, we impose controlled missing data scenarios to a potential effect modifier from the 
Work, Family, and Health Study. 
%to further compare the available methods using real data. 
Our results suggest that multilevel multiple imputation (MMI) and Bayesian MMI have better performance than other available methods, and that Bayesian MMI has lower bias and closer to nominal coverage than standard MMI when there are model specification or compatibility issues. %We also provide recommendations for practitioners and outline future research areas.
\end{abstract}


\maketitle

\section{Introduction}
Cluster-randomized trials (CRTs) are a popular experimental design where groups of individuals are randomized to different treatment arms.\cite{murray1998} This design is used when the intervention is designed to be administered at the cluster level or in studies where randomization of individuals is difficult or impractical. Though methods for study design and analysis of CRTs have been developed (see, for example, a pair of recent reviews by Turner et al.\cite{turner2017design,turner2017analysis}), they mostly focus on studying the average treatment effect. Methods for the design and analysis of CRTs have only recently branched into identifying heterogeneous treatment effects (HTE), where the treatment effect is hypothesized to vary across pre-specified subgroups of the trial population.\cite{yang2020,tong2022,li2022designing,tong2023sample,ryan2023maximin,maleyeffsample} In particular, there is burgeoning interest in confirmatory HTE analysis of CRTs focused on pre-specified subgroups often defined by baseline demographic characteristics. A recent systematic review of 64 CRT analyses published between 2010 and 2016 found that 16 (25\%) examined HTE across pre-specified demographic subgroups, and noted that guidance on design and analysis of CRTs to assess HTE needs further development.\cite{starks2019} Evaluation of HTE is critical to understanding treatment effects in vulnerable subpopulations and developing real-world recommendations for the administration of new treatments.

In this article, we focus on confirmatory and hypothesis-driven HTE analyses that require pre-specification. This is in contrast to exploratory HTE analyses that are often \emph{ad-hoc} and used for generating hypotheses for future studies. \textcolor{blue}{Confirmatory HTE analyses in CRTs can be assessed using a variety of methods, including GEE or generalized mixed-effects models with statistical interaction terms,\cite{yang2020} pre-specified subgroup analyses (often also based on GEE or mixed-effects models),\cite{wang2023sample} and data-driven methods for targeting pre-specified conditional average treatment effects for well-defined subpopulations.\cite{abrevaya2015,semenova2021} Among these approaches, the statistical interaction approach is likely the most popular in practice.\cite{starks2019,nicholls2023health}} These interactions are often assessed one at a time, but the potential effect modifiers composing the interactions may be subject to missingness. \textcolor{blue}{As these modifiers are typically baseline covariates, they can be missing in a variety of common scenarios, including when electronic health record data is used to capture baseline covariates or when trial participants actively choose not to respond to questions in baseline surveys and enrollment forms. This latter scenario may occur if the questions make participants uncomfortable, if participants are not sure how to answer the questions, or if there are language barriers, among other reasons.} Missing data \textcolor{blue}{of various forms} can introduce bias in the analysis and interpretation of study results,\cite{little2019} and missing modifiers in particular have been shown to introduce bias in the estimation of interaction terms in observational studies and individually-randomized trials.\cite{von2009,seaman2012} However, the extent of this bias and appropriate methods to address missing effect modifier data are under-explored in CRTs. \textcolor{blue}{In the remainder of this section, we review some existing methods from two related areas of research: missing outcomes in CRTs and missing modifiers in individually-randomized trials and observational studies, to provide context for our current study.}

While research on missing effect modifier data in CRTs is under-developed, research on methods for missing outcome data in CRTs has grown substantially over the last two decades.\cite{diaz2014missing,fiero2016,turner2017analysis} In particular, multiple imputation (MI) has been shown to be a useful tool for reducing bias,\cite{hunsberger2001} but addressing the correlations within clusters was found to be key for valid estimation and inference in CRTs.\cite{taljaard2008} One method that addresses such correlations is multilevel multiple imputation (MMI), which includes random intercepts for clusters in both imputation and outcome models. MMI has been studied for both continuous and binary outcomes in CRTs,\cite{caille2016,hossain2017binary,turner2020} and has also been shown to be useful for studies with missing multivariate outcomes.\cite{diaz2014handling} While fixed effects for clusters are used in some imputation models, such specification may introduce model parameter estimation bias in many settings, especially when the ICCs or cluster sizes are small.\cite{andridge2011} Whether and how these methods can be applied to study HTE in CRTs with missing effect modifier data is comparatively less well understood; the goal of this article is therefore to compare existing methods such as MI and MMI in a CRT setting with a continuous outcome where a binary effect modifier (used to define pre-specified subgroup membership) is subject to missingness.

Research developing or comparing methods for addressing missing modifier data is more developed in the observational study and individually-randomized trial settings. \textcolor{blue}{In some of this existing literature, these methods are described broadly as methods for missing covariates or missing covariates with interactions. It is worth mentioning that in the current article, all effect modifiers of interest we consider are baseline covariates, even though not all baseline covariates are effect modifiers of interest.} In observational studies, MI and MMI are popular for handling missing modifiers, and much attention has been focused on whether interaction terms should be computed as the product of variables after imputation (passive imputation) or treated as their own missing variable in the process (just-another-variable imputation).\cite{seaman2012} This issue is less important in randomized trials focused on HTE, since an interaction between treatment and a modifier will always equal 0 under placebo and equal the modifier under treatment (assuming 0-1 coding). A second key issue, which remains relevant for randomized trials, is that imputation and outcome models should be compatible. Imputation and outcome models are compatible (sometimes referred to as congenial) if there exists a joint model for the modifier and outcome which has implied conditional distributions equal to those specified by the imputation and outcome models.\cite{meng1994,bartlett2015} However, choosing a compatible imputation model may not be straightforward, especially when the outcome model contains interaction terms. \textcolor{blue}{Semiparametric approaches such as fractional imputation\cite{yang2017,sang2022} and nonparametric approaches such as imputation from Bayesian Additive Regression Trees\cite{xu2016,ramosaj2019} have been proposed to address missing covariates, but such methods have not yet been extended to account for the within-cluster correlations in the covariates and outcomes in the CRT setting.} 
%are not readily adaptable to CRTs.} 
With independent data, Erler et al.\cite{erler2016} noted that nesting MI in a Bayesian approach could alleviate imputation model misspecification issues and help account for additional uncertainty in the imputation process. This procedure can be adapted for the analyses of CRTs with a random-effects specification in the imputation model to arrive at a Bayesian MMI (B-MMI) approach, which represents another candidate method which may be useful for evaluating HTE in CRTs with missing modifiers. \textcolor{blue}{Thus, from prior work on (i) missing outcomes in CRTs and (ii) missing modifiers in individually-randomized trials or observational studies, we can enlist a set of existing methods that may be applicable for assessing HTE in CRTs with missing modifiers, but which have not been studied specifically for that purpose. Importantly, previous methods and guidelines developed for individually-randomized trials will not necessarily be applicable for CRTs because covariates and/or missingness may be correlated within clusters, and this challenge has motivated our current study.} 

Overall, the contribution of this article is to develop and report on a series of comparison studies in order to determine which readily available methods may be most suitable for assessing HTE in CRTs with missing effect modifier data. The rest of the article is structured as follows. First, we define notation and describe the generalized estimating equations (GEEs),\cite{liang1986} which will be used for all outcome models throughout the article for final outcome analyses. Then, we describe each of the missing data methods that will be compared. Next, we compare the methods in an extensive simulation study using the ADEMP framework.\cite{morris2019} Then, we apply each of the methods to real data from the Work, Family, and Health Study, where we impose controlled missing data scenarios on complete data where the true values are known. Finally, we conclude with a discussion of the results and offer some recommendations for practitioners.

\section{Notation and Outcome Model for Cluster-Randomized Trials}

We consider the setting of a two-arm CRT with intervention assignment for cluster $i$ denoted as $A_{i} = 1$ for treatment and $A_{i} = 0$ for control. Suppose there are $C$ clusters of potentially varying sizes $n_{i}$ and let the total sample size $N = \sum_{i=1}^{C} n_{i}$. Let $M_{ij}$ be a pre-specified effect modifier (or subgroup variable) of interest for individual $j$ in cluster $i$ and let $Y_{ij}$ be the observed outcome for individual $j$ in cluster $i$. \textcolor{blue}{In this article, we will focus specifically on a binary effect modifier measured at the individual level and a continuous outcome variable.} Let $R_{ij}$ be an indicator equal to 1 if $M_{ij}$ is measured and 0 if $M_{ij}$ is missing. Suppose that there are $p$ additional baseline covariates available denoted $\bcx_{ij} = (X_{1ij}, X_{2ij}, \ldots, X_{pij})^{T}$. We assume $\bcx_{ij}$ are fully observed auxiliary variables that may be marginally or conditionally associated with the missing effect modifier.

%Each of the comparison missing data methods 
We first describe the data and analytic model in cases where the effect modifier is fully observed. \textcolor{blue}{We will focus on a GEE approach with correctly specified mean model for testing HTE to streamline our discussion. We consider the GEE approach because (i) the estimation and inference for treatment effect parameters can be robust to misspecification of the within-cluster correlation structure under correct specification of the marginal mean model,\cite{liang1986} and (ii) the population-averaged interpretation of the marginal mean model may be preferable for the analysis of CRTs.\cite{preisser2003integrated} We acknowledge, however, that GEE is not the only possible outcome modeling choice for CRTs, and mixed-effects regression may provide a more efficient estimator when both the conditional mean model and the random-effects structure are correctly specified.} 
%only possible outcome modeling choice for CRTs, but is chosen to help streamline the article by considering one general substantive model form and estimation procedure throughout. It also will allow for straightforward targeting of marginal estimands when comparing methods.} 
Suppose that the outcome follows a generalized linear marginal model of the form
\begin{equation*}
    g\{ E(Y_{ij}) \} = g(\mu_{ij}) = \gamma_{0} + \gamma_{1}A_{i} + \gamma_{2}M_{ij} + \gamma_{3}A_{i}M_{ij},
\end{equation*}
where $g$ is a link function, and $\mu_{ij}$ is the mean function of individual $j$ in cluster $i$. GEE methods estimate the model parameters $\bfgamma = (\gamma_{0}, \gamma_{1}, \gamma_{2}, \gamma_{3})^{T}$ while allowing for correlation between the model outcomes across individuals. This is accomplished by specifying the structure of a working covariance matrix $\bcv_{i}$ for $\bcy_{i} = (Y_{i1}, Y_{i2}, \ldots, Y_{in_{i}})^{T}$ and solving the estimating equations
\begin{equation*}
    \sum_{i=1}^{C} \frac{\partial \bmu_{i}}{\partial \bfgamma^T} \bcv_{i}^{-1} (\bcy_{i} - \bmu_{i}) = 0,
\end{equation*}
where $\bmu_{i} = (\mu_{i1}, \mu_{i2}, \ldots, \mu_{in_{i}})^{T}$. Common specifications of the working covariance structure include independence or exchangeability of individuals within a cluster. Importantly, when the mean model is correctly specified, the GEE estimator remains consistent regardless of covariance specification,\cite{zeger1986} but can usually be more efficient when the working covariance structure is correctly specified.\cite{liang1986,li2021sample-bin} In CRTs, an exchangeable correlation matrix is commonly specified to acknowledge the correlated outcomes in each cluster. Typically, robust sandwich standard error estimators are used for inference, and remain consistent estimators even if the working covariance model is incorrectly specified. The sandwich variance estimator takes the form of
\begin{equation*}
    \widehat{\bdelta} \left\{\sum_{i=1}^{C} \frac{\partial \bmu_{i}}{\partial \bfgamma^T} \bcv_{i}^{-1} (\bcy_{i} - \widehat{\bmu}_{i}) (\bcy_{i} - \widehat{\bmu}_{i})^{T} \bcv_{i}^{-1} \frac{\partial \bmu_{i}}{\partial \bfgamma} \right\}\widehat{\bdelta} ,
\end{equation*}
where $\widehat{\bdelta} = \sum_{i=1}^{C} \{ (\partial \bmu_{i} / \partial \bfgamma^T) \bcv_{i}^{-1} (\partial \bmu_{i} / \partial \bfgamma) \}^{-1}$ is referred to as the model-based variance estimator. \textcolor{blue}{Of note, even though the cluster sizes are potentially varying, we have assumed the absence of informative cluster size\cite{kahan2023estimands} and a correct specification of the GEE marginal mean model. In this case, we can define the two subgroup-specific treatment effect estimands on the link function scale by $\gamma_1=g\{ E(Y_{ij}|A_i=1,M_{ij}=0) \}-g\{ E(Y_{ij}|A_i=0,M_{ij}=0) \}$ and $\gamma_1+\gamma_3=g\{ E(Y_{ij}|A_i=1,M_{ij}=1) \}-g\{ E(Y_{ij}|A_i=0,M_{ij}=1) \}$; therefore, $\gamma_1+\gamma_3E(M_{ij})$ measures the average treatment effect, whereas $\gamma_3$ represents the difference between the two subgroup-specific treatment effects and measures the degree of treatment effect heterogeneity.} 
%In the absence of informative cluster size, the average treatment effect parameter and the subgroup-specific treatment effect parameters may be interpreted as either a cluster-level estimand and an individual-level estimand; see, for example, the discussion in Section 2.2 of Wang et al.\cite{wang2021mixed}}
%Note that this outcome model will target a participant-level effect rather than a cluster-level effect; this convention is followed throughout the article, but similar results should apply to models that target cluster-level treatment effects.}

Next, five missing data methods associated with fitting the GEE outcome model when $M_{ij}$ is partially missing are described in detail. We will develop a simulation study and application to compare these approaches to generate practical recommendations for analyzing CRTs. We specifically consider two types of missing data mechanisms. First, when $R_{ij} \perp \!\!\! \perp M_{ij}$, the effect modifier data are missing completely at random (MCAR). When the modifier missingness depends only on observed variables, i.e., $R_{ij} \perp \!\!\! \perp M_{ij} | A_{i}, Y_{ij}, \bcx_{ij}$, the data are said to be missing at random (MAR). When the modifier missingness depends on the values of the modifier or other unobserved data, the data are considered missing not at random (MNAR); we will not address MNAR scenarios in our simulations or application, but will return to a discussion of this at the end. %Within the scope of this article,  will not be considered.

\section{Statistical Methods for Addressing a Missing Binary Effect Modifier in CRTs}

\subsection{Complete-Case Analysis}

Complete-case analysis (CCA) entails fitting a model based on only individuals for whom complete data is available. Thus, any individuals with missing effect modifiers will be excluded from the analysis data set. While this is usually a valid strategy in MCAR settings, it may lead to biased inference in MAR settings where the outcome and missingness of the modifier are both dependent on one or more baseline covariates. In addition, CCA may have lower power than other methods which utilize the full data. Performing CCA is equivalent to solving the GEE estimating equations weighted by the indicator that the modifier is observed,
\begin{equation*}
    \sum_{i=1}^{C} \frac{\partial \bmu_{i}}{\partial \bfgamma^T} \bcr_{i} \bcv_{i}^{-1} (\bcy_{i} - \bmu_{i}) = 0,
\end{equation*}
where $\bcr_{i}$ is a diagonal matrix with diagonal vector $\{R_{i1}, R_{i2}, \ldots, R_{in_{i}}\}^{T}$. The robust sandwich variance estimator can be derived analogously by including the missing indicator matrix $\bcr_i$. 

\subsection{Single Imputation}

\textcolor{blue}{When the data are not MCAR, it is often reasonable to relax this assumption and assume the data are MAR. This assumption underlies each of the imputation methods described in this article.} Single imputation (SI) entails imputing a single, fixed value for each of the missing effect modifiers and then fitting the GEE outcome model. Imputation can be achieved in many ways, but is typically performed using parametric models. As stated in the Introduction, the distribution of $M_{ij} | Y_{ij}, A_{i}, \bcx_{ij}$ implied by the outcome model and other assumptions may not correspond to a model fit by off-the-shelf software. However, parametric models may still be ``approximately compatible'' in many scenarios, and it is recommended in general that the imputation model contain interaction terms that correspond to those in the outcome model (e.g., if the outcome model contains an interaction between the treatment and modifier, then the imputation model should contain an interaction between the outcome and treatment).\cite{tilling2016}

Once the imputation model is specified and fit, missing values of $M_{ij}$ are replaced by predicted values from the model. Let $M_{ij}^{*}$ be equal to $M_{ij}$ whenever $R_{ij} = 1$ and equal to the imputed value whenever $R_{ij} = 0$, and let $g(\mu_{ij}^{*}) = \gamma_{0} + \gamma_{1}A_{i} + \gamma_{2}M_{ij}^{*} + \gamma_{3}A_{i}M_{ij}^{*}$ be the specified mean model for the GEE analysis with imputed, complete data. Then the outcome model parameters are estimated using the imputed data set by solving the estimating equations
\begin{equation*}
    \sum_{i=1}^{C} \frac{\partial \bmu_{i}^{*}}{\partial \bfgamma^T} \bcv_{i}^{-1} (\bcy_{i} - \bmu_{i}^{*}) = 0.
\end{equation*}

\subsection{Multiple Imputation}\label{sec:MI}

MI extends the idea of single imputation by imputing multiple values for each missing effect modifier and then combining results across imputed data sets in order to account for variability in the imputation procedure. In particular, $D$ unique data sets will be imputed, where $D \in [5, 15]$ is often recommended in practice.\cite{schafer1997} In the simulations and data application of this article, we will use $D = 15$. Then the outcome model will be fit for each of the $D$ data sets and the model estimates will be combined using Rubin's rule.\cite{rubin1996,rubin2004} For example, let $\widehat{\bfgamma}_{d}$ be the estimate of $\bfgamma$ from imputed data set $d$ for $d = 1,2,\ldots,D$. Then $\widehat{\bfgamma}^{MI} = \frac{1}{D}\sum_{d=1}^{D} \widehat{\bfgamma}_{d}$ and
\begin{align*}
    \widehat{\text{Var}}(\widehat{\bfgamma}^{MI}) &= \frac{1}{D} \sum_{d=1}^{D} \text{Var}(\widehat{\bfgamma}_{d}) + \frac{D+1}{D} \cdot \frac{1}{D-1}\sum_{d=1}^{D} (\widehat{\bfgamma}_{d} - \widehat{\bfgamma}^{MI})^{2}.
\end{align*}
Standard confidence intervals can be constructed for $\bfgamma$ noting that $(\widehat{\bfgamma}^{MI} - \bfgamma) / \sqrt{\widehat{\text{Var}}(\widehat{\bfgamma}^{MI})} \sim t_{\nu}$. The degrees of freedom $\nu$ are calculated as
\begin{equation*}
    \nu = \left(D - 1\right) \left(1 + \frac{\sum_{d=1}^{D} \text{Var}(\widehat{\bfgamma}_{d})}{(D + 1)\sum_{d=1}^{D} (\widehat{\bfgamma}_{d} - \widehat{\bfgamma}^{MI})^{2} / (D - 1)} \right)^{2}
\end{equation*}

When the number of clusters are limited, an adjusted degrees of freedom $\nu_{adj}$ has been recommended in practice for CRTs.\cite{barnard1999,taljaard2008} This is calculated as
\begin{equation*}
    \nu_{adj} = \left( \frac{1}{\nu} + \frac{1}{\nu_{obs}} \right)^{-1},
\end{equation*}
where
\begin{equation*}
    \nu_{obs} = \frac{(C - 1)(C - 2)}{C + 1} \left( 1 + \frac{(D + 1)\sum_{d=1}^{D} (\widehat{\bfgamma}_{d} - \widehat{\bfgamma}^{MI})^{2}}{(D - 1)\sum_{d=1}^{D} \widehat{\bfgamma}_{d}} \right)^{-1}
\end{equation*}
These adjusted degrees of freedom are used for all MI and MMI methods throughout the article.

\subsection{Multilevel Multiple Imputation}

While the above imputation approaches each use GEE to account for intracluster correlation (ICC) in the outcome variable when analyzing complete data, they assume in the imputation process that $\text{ICC} = 0$ for the effect modifier. This may be overly simplistic in CRTs where the effect modifiers (and covariates in general) in the same cluster can be positively correlated, leading to a non-zero covariate ICC.\cite{raudenbush1997statistical,yang2020,tong2022,maleyeffsample} Ignoring the covariate ICC in the imputation process may lead to incorrect confidence intervals around the HTE estimator. In the context of non-zero covariate ICC, MMI entails following the multiple imputation procedure described above, but using a multilevel imputation model that acknowledges the correlated nature of the effect modifiers. For example, instead of imputing a binary effect modifier using logistic regression, one might use a logistic linear mixed-effects model with a random intercept corresponding to cluster membership, where the empirical best linear unbiased predictions can be used to predict the missing effect modifier. Then the imputed data sets are combined using Rubin's rule, as above. The MI and MMI imputation procedures are implemented by predicting from models in a Frequentist fashion, i.e., treating the estimated imputation model parameters as fixed. This is sometimes referred to as improper MI or approximate MI.\cite{caille2016,murray2018}

\subsection{Bayesian Multilevel Multiple Imputation}

B-MMI implements MMI within a Bayesian framework to account for uncertainty in imputation model parameter estimation. This can be accomplished using Markov Chain Monte Carlo methods such as Gibbs sampling.\cite{casella1992} For a fully Bayesian approach that integrates the imputation model and outcome model in a single procedure, the general algorithm can be summarized as follows: (i) specify priors for all parameters; (ii) draw posterior samples of the imputation model parameters and impute values for all missing modifiers; (iii) draw posterior samples of the outcome model parameters using the imputed data; (iv) iterate between steps (ii) and (iii) until convergence, and summarize the posterior distribution of the outcome model parameter estimator. Although this fully Bayesian approach is attractive, it often does not separate the imputation step and the outcome data analysis step, and may constrain the flexibility on choice of outcome model for data analysis. In practice, to separate the imputation and outcome data analysis steps, one often implements Bayesian multiple imputation by only omitting step (iii), and then sampling $D$ complete data sets from the posterior distribution, perhaps using a thinning procedure. Once the complete data are obtained, then one can fit the outcome data analysis model (e.g., the GEE procedure) and combine the $D$ final parameter estimates using Rubin's rule. 

%then sample $D$ parameter estimates, perhaps using a thinning procedure; (v) Combine final parameter estimates acquired in Step 4 using the procedures described for MI and MMI;
% \begin{enumerate}
%     \item Specify priors for all parameters;
%     \item Draw posterior samples of the imputation model parameters and impute values for all missing modifiers;
%     \item Draw posterior samples of the outcome model parameters using the imputed data;
%     \item Iterate between steps 2 and 3 until convergence, then sample $D$ parameter estimates, perhaps using a thinning procedure;
%     \item Combine final parameter estimates acquired in Step 4 using the procedures described for MI and MMI;
% \end{enumerate}

In this article, scenarios with a binary effect modifier will be considered. A natural choice for an imputation model would be a logistic mixed-effects model, with a random intercept corresponding to cluster membership. Leveraging P\'{o}lya-Gamma random variables for Bayesian inference of logistic regression models was proposed by Polson et al.\cite{polson2013}, who also described expanding the procedure to logistic mixed-effects models. \textcolor{blue}{The essence of this approach is to recognize that the binomial likelihood parameterized by log-odds can be written as a Gaussian mixture of P\'{o}lya-Gamma densities, namely
\begin{align*}
\frac{(e^{\psi})^a}{(1+e^{\psi})^b}=2^{-b}e^{(a-b/2)\psi}\int_0^{\infty}e^{-\omega\psi^2/2}f(\omega|b,0)d\omega,
\end{align*}
for all real numbers $a$, where $b>0$, $f(\omega|b,0)$ is the density of the P\'{o}lya-Gamma density with parameters set to be $b$ and $0$, or $PG(b,0)$, and $\psi$ can be taken as the linear component in a logistic mixed-effects model.} However, handling missing data via this approach has not been described to our knowledge. To formally describe such a procedure, we combine the aforementioned generic approach with standard techniques for handling missing data in a Bayesian procedure using the following proposed Gibbs sampler. \textcolor{blue}{Specifically, we write the imputation model as $\text{logit} \{ P(M_{ij} = 1) \} = \bw_{ij}^T\bfeta + \alpha_{i}$, where $\bw_{ij}$ is the $p \times 1$ design vector including auxiliary covariates and possibly the treatment and observed outcome, and $\alpha_{i}\sim N(0,1/\tau_\alpha^2)$ is the cluster-level random intercept, with $\tau_\alpha^2$ defined as the precision parameter.} Let $\widehat{\bfeta}_{O}$ be the estimates from a logistic mixed-effects imputation model fit to the observed data (with initialized missing modifier values). Then,
\begin{enumerate}
    \item Set initial values for the imputation model parameters $\bfeta$ as $\widehat{\bfeta}_{O}$.
    \item Set initial values for random intercepts $\balpha=(\alpha_{1},\ldots,\alpha_{C})^T$ to 0 and an initial random effect precision $\tau_{\alpha} = 0.5$.
    \item Define a prior $p(\bfeta) \sim N(\widehat{\bfeta}_{O}, \bsigma)$ where $\bsigma$ is a $p \times p$ diagonal matrix with each variance set to be a large value, such that the prior is uninformative. Set Gamma hyperpriors for the random effect precision to be $c = d = 0.01$.
    \item Impute missing values of $M_{ij}$ using the latest iteration of imputation model parameters.
    \item Update the imputation model parameters and random effects as follows.
    \begin{enumerate}
        \item Let $\bcw=(\bw_{1},\ldots,\bw_{C})^T$ be the full design matrix for the imputation model and $\bcm$ be the full modifier vector $(\bcm_{1},\ldots,\bcm_{C})^{T}$, where $\bw_i=(\bw_{i1},\ldots,\bw_{in_i})^T$ and
        $\bcm_{i} = (M_{i1}, M_{i2}, \ldots, M_{in_{i}})$. Let $\bci_{C}$ be an $N \times C$ matrix corresponding to cluster membership such that each element of column $i$ is 1 for all individuals that belong to cluster $i$ and 0 otherwise.
        \item Generate $\bfomega$ as a Polya-Gamma random vector $\bfomega \sim PG(1, \bcw^{T}\bfeta + \bci_{C}\balpha$). Let $\bomega$ be a diagonal matrix with the diagonal vector equal to $\bfomega$.
        \item Update $\bfeta$ as a multivariate random Normal variable with:
        \begin{enumerate}
            \item Covariance parameter equal to $(\bcw \bomega\bcw^{T} + \bsigma^{-1})^{-1}$.
            \item Mean parameter equal to $(\bcw \bomega \bcw^{T} + \bsigma^{-1} )^{-1} \cdot [\bsigma^{-1}\widehat{\bfeta}_{O} + \bcw \bomega \{ (\bcm - 0.5) \circ \bfomega^{-1} - \bci_{C}\balpha \}]$, where $\circ$ is an element-wise product.
        \end{enumerate}
        \item Update random effects $\balpha$ as a multivariate random Normal variable with:
        \begin{enumerate}
            \item Covariance parameter equal to $(\tau_{\alpha} + \bci_{C}^{T} \bomega \bci_{C})^{-1}$.
            \item Mean parameter equal to $ (\tau_{\alpha} + \bci_{C}^{T} \bomega \bci_{C})^{-1} \cdot \bci_{C}^{T} \cdot \{\bcm - 1/2 - \bcw^{T} \bfeta \circ \bfomega\}$.
        \end{enumerate}
        \item Generate $\tau_{\alpha}$ as a Gamma random variable $\tau_{\alpha} \sim \text{Gamma}(1, c + C/2, d + (\bci_{C}\balpha)^{T}\bci_{C}\balpha/2)$.
    \end{enumerate}
    \item Iterate between Steps 4 and 5 until desired MCMC chain length is reached (or parameter estimates have sufficiently converged).
\end{enumerate}

Finally, one can fit the GEE outcome model on each of $D$ imputed data sets and combine the parameter estimates using Rubin's rule. One can select the $D$ data sets after a burn-in period and use a thinning procedure to separate imputed data sets across the MCMC chain. Note that this method is not fully Bayesian, in the sense that the imputation and outcome models are not estimated in the same procedure under a combined joint likelihood. However, the method does account for uncertainty in estimation of the imputation model parameters, distinguishing it from the MMI approach. \textcolor{blue}{Although the exact performance of this specific procedure is unknown in our setting and will be explored in the next section, prior work on multiple imputation has demonstrated that Bayesian imputation procedures should generally result in improved performance over their Frequentist counterparts (approximate MI procedures) from a theoretical perspective. Moreover, in scenarios where the imputation and outcome models may not be necessarily compatible, Bayesian imputation procedures hold the promise to explicitly account for uncertainty in estimating the imputation model parameters.\cite{meng1994,zhang2003,murray2018}}

\section{Simulation Study}

\subsection{Aims}

The aim of the simulation study is to compare the missing data methods described in the Methods section under plausible data-generating mechanisms for CRTs. The key study objectives are to (i) study which of the methods perform best in practice when imputation models are correctly specified and (ii) evaluate how robust each approach is to imputation model misspecification and/or lack of compatibility between the imputation and outcome models. We will evaluate these questions under several specific data-generating processes described below.

\subsection{Data-generating Mechanisms}

Simulations were separated into two scenarios according to the data-generating mechanisms. The initial setup was identical in each scenario. First, a set of $C \in \{ 20, 50, 100 \}$ clusters were generated where cluster $i$ had $n_{i}$ individuals, and $n_i$ is sampled from the Poisson distribution with mean 50. The total sample size was $N = \sum_{i=1}^{C} n_{i}$. Then, a binary treatment $A_{i}$ was randomized at the cluster level, with $P(A_{i} = 1) = 0.5$ and exact 1:1 allocation. Next, a binary effect modifier $M^{\dagger}_{ij}$ was generated as a Bernoulli random variable with $\text{logit} \{ P(M^{\dagger}_{ij} = 1) \} = 0.5 + \alpha_{i}$ where $\alpha_{i}$ was a random intercept generated as a Normal random variable with mean 0 and variance such that the ICC defined on the latent response scale was equal to 0.1.\cite{eldridge2009intra,li2017evaluation,maleyeffsample} Subsequent data-generating mechanisms varied across the two scenarios.

\textbf{Scenario 1:} In the first scenario, an additional covariate for person $j$ in cluster $i$, $X_{ij}$, was generated as a standard Normal random variable. Then the outcome $Y_{ij}$ was simulated according to the model
\begin{align*}
    Y_{ij} = 1 &+ 1A_{i} + 0.75M^{\dagger}_{ij} + \beta_{3}A_{ij}M^{\dagger}_{ij} + 0.8X_{ij}A_{i} - 0.4X_{ij}M^{\dagger}_{ij} + 0.7X_{ij}A_{i}M^{\dagger}_{ij} + \kappa_{i} + \epsilon_{ij}
\end{align*}
where $\kappa_{i}$ was a random intercept generated as a Normal random variable with a mean of 0 such that the outcome ICC was equal to 0.1 and $\epsilon$ was a Normal random variable with a mean of 0 and variance equal to 3. The outcome was simulated under two possible coefficient values, $\beta_{3} = 0$ and $\beta_{3} = -\{1 + \exp(-0.5)\}$. Finally, missingness was imposed on the effect modifier by generating an indicator $R_{ij}$ as a Bernoulli random variable with $\text{logit} \{ P(R_{ij} = 1) \} = 1.2 + 0.5X_{ij} - 0.2Y_{ij}$ such that the observed $M_{ij} = M^{\dagger}_{ij}$ when $R_{ij} = 1$ and is missing otherwise. While outcome-dependent missingness may seem unusual for a clinical trial setting with outcome measured after treatment and modifier measurement, this effectively captures a scenario where unmeasured covariates affect both missingness and the outcome. \textcolor{blue}{The marginal percentage of missingness was about 32\% when $\beta_{3} = 0$ and 30\% when $\beta_{3} = -\{1 + \exp(-0.5)\}$.}

\textbf{Scenario 2:} In the second scenario, three covariates $X_{1ij}, X_{2ij},$ and $X_{3ij}$ were generated as independent standard Normal random variables. Then the outcome was simulated according to the model
\begin{align*}
    Y_{ij} = 1 &+ 1A_{i} + 0.75M^{\dagger}_{ij} + \beta_{3}A_{ij}M^{\dagger}_{ij} + 0.8X_{1ij}A_{i} - 0.4X_{1ij}M^{\dagger}_{ij} \\
    &+ 0.7X_{1ij}A_{i}M^{\dagger}_{ij} + 0.9X_{2ij}A_{i}M^{\dagger}_{ij} - 1.1X_{3ij}A_{i}M^{\dagger}_{ij} + \kappa_{i} + \epsilon_{ij}
\end{align*}
where $\kappa_{i}$, $\epsilon$, and $\beta_{3}$ were generated as in Scenario 1. Missingness was imposed on the effect modifier by generating an indicator $R_{ij}$ as a Bernoulli random variable with $\text{logit} \{ P(R_{ij} = 1) \} = 1.5 + 0.6X_{1ij} + 1.2X_{2ij} - 0.8X_{3ij} - 0.2Y_{ij} + \zeta_{i}$ such that the observed $M_{ij} = M^{\dagger}_{ij}$ when $R_{ij} = 1$ and is missing otherwise. The random intercept $\zeta_{i}$ was generated with mean 0 such that the missingness ICC on the latent response scale was equal to 0.1.\cite{eldridge2009intra} \textcolor{blue}{The marginal missingness percentage was the same as Scenario 1.} Directed acyclic graphs representing Scenarios 1 and 2 are provided in Figure 1(a) and 1(b), respectively. All data simulation and analyses were performed in R Software Version 4.1.2.\cite{r2021} Data were generated using the 64-bit Mersenne-Twister with input seed $1000k$ for simulation iteration $k$ in each scenario.

Note that in each scenario, the modifier is generated independently of the baseline covariates. This may seem unusual given the stated goal to compare imputation methods. However, this choice helps illustrate model compatibility in this setting. The modifier $M$ is associated with the outcome $Y$, such that $Y$ should in principle be included in imputation model specifications. However, even though $M$ is marginally independent of $A$, $X_{1}, X_{2}$, and $X_{3}$, once we condition on $Y$, it is not conditionally independent of those variables. Thus, the outcome model generation process informs a fairly complex imputation model, despite the simple modifier generation process.

\begin{figure*}[htb]
\begin{center}
\fbox{\includegraphics[width=5.5in, trim={4.5cm 11.5cm 3cm 12cm}]{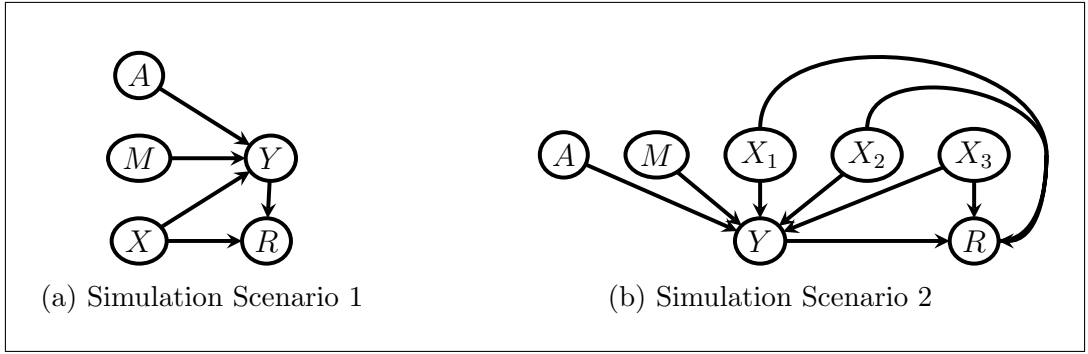}}
\caption{Directed acyclic graphs (DAGs) corresponding to the simulation scenarios considered.}
\end{center}
\end{figure*}

\subsection{Estimands}

Two estimands were considered for each simulation study. First, we consider the implied marginal model after integrating out additional covariates, e.g., for the first scenario,
%\begin{align*}
    %E(Y_{ij}|A_i,M^{\dagger}_{ij}) = 1 &+ 1A_{i} + 0.75M^{\dagger}_{ij} + \beta_{3}A_{ij}M^{\dagger}_{ij} \\
     %&+ 0.8E(X_{ij})A_{i} - 0.4E(X_{ij})M^{\dagger}_{ij} \\
     %&+ 0.7E(X_{ij})A_{i}M^{\dagger}_{ij} \\
    %= \;\;\; &\gamma_{0} + \gamma_{1}A_{i} + \gamma_{2}M^{\dagger}_{ij} + \gamma_{3}A_{i}M^{\dagger}_{ij}
%\end{align*}
\begin{align*}
    E(Y_{ij}|A_i,M^{\dagger}_{ij}) &= 1 + A_{i} + 0.75M^{\dagger}_{ij} + \beta_{3}A_{ij}M^{\dagger}_{ij} + 0.8E(X_{ij})A_{i} - 0.4E(X_{ij})M^{\dagger}_{ij} + 0.7E(X_{ij})A_{i}M^{\dagger}_{ij} \\
    &= \gamma_{0} + \gamma_{1}A_{i} + \gamma_{2}M^{\dagger}_{ij} + \gamma_{3}A_{i}M^{\dagger}_{ij}
\end{align*}
The first estimand of interest is the interaction term from the marginal model, $\gamma_{3}$, whose interpretation as the difference in subgroup-specific treatment effects is provided earlier. This interaction term will be referred to as the HTE estimand moving forward. Note that since all baseline covariates had mean 0, the HTE estimand is equal to $\beta_{3}$ in both scenarios. The second estimand of interest is the average treatment effect (ATE), namely, $E(Y | A = 1) - E(Y | A = 0)$. As we mentioned earlier, this estimand is equal to $\gamma_{1} + \gamma_{3}E(M^{\dagger}) = 1 - \beta_{3}E(M^{\dagger})$ under both scenarios since $X, A,$ and $M^{\dagger}$ are marginally independent under the full data distribution. Thus, the true ATE estimand is 1 when $\beta_{3} = 0$ and the true ATE estimand is 0 when $\beta_{3} = -\{1 + \exp(-0.5)\}$.

\subsection{Methods}

Each of the missing data methods described earlier were used to analyze the data generated under each scenario. In particular, data were analyzed using CCA, SI, MI, MMI, and B-MMI. For each method, the final outcome data analysis model was a correct GEE specification for the marginal mean model (integrating out baseline covariates) with the identity link function. Thus the estimator of the HTE estimand $\widehat{\gamma}_{3}$ is simply given by the estimated interaction coefficient from each model. For each analysis, the corresponding robust standard error is used. The estimator for the ATE estimand is given by $$\widehat{\gamma}_{1} + \widehat{\gamma}_{3}\sum_{i=1}^{C}\sum_{j=1}^{n_{i}} M^{*}_{ij} / N,$$ 
where $M^{*}_{ij} = M_{ij}$ when the modifier is observed and equals the imputed values otherwise. More simply, one can mean-center the modifier variable after imputation such that the estimator is given by $\widehat{\gamma}_{1}$.\cite{tong2022,li2022designing} The standard error of the ATE estimator can then also be estimated directly as the robust standard error of the treatment coefficient, or can be obtained via the Delta Method if the modifier is not mean-centered. Mean-centering of the effect modifier, however, does not affect the interpretation of the interaction effect or HTE estimator. 

\textcolor{blue}{The general imputation model used had the form $$\text{logit}\{ P(M_{ij} = 1) \} = \eta^{T}f(X_{1ij}, X_{2ij},X_{3ij}, A_{i}, Y_{ij}) + \xi_{i},$$ 
where the function $f$ returns a design vector that is a function of its arguments and is varied by model specification; the random intercept $\xi_{i}$ was only included for multilevel multiple imputation methods. Even though we have considered continuous baseline covariates in the simulations, the general form of the imputation model can easily include binary and categorical predictors as in any conventional regression model (and indeed includes the binary treatment assignment).} For each of the imputation-based approaches, 5 imputation model specifications were used, ranging from models that only contained main effects (likely highly incompatible with the outcome model) to ones with three-way interactions and lower order terms (approximately compatible with the outcome model). Table 1 elaborates on the details of the imputation model specifications. Overall, 2 data scenarios were compared across 3 sample sizes for 2 estimands, 2 choices of $\beta_{3}$ values, and 21 combinations of methods and model specifications, for a total of 504 comparisons.

%\begin{landscape}
\begin{sidewaystable*}[]
    \begin{adjustwidth}{-2.65cm}{-1cm}
    \centering
    \footnotesize
    \caption{Possible imputation model specifications considered in the simulation studies, as well as abbreviations used in communicating simulation results. Specification 1 (2) is the model used for simulation scenario 1 (2). For MMI and B-MMI methods, a random intercept was added to each of the specifications in the table. $\text{logit}(\cdot) = \text{log}(\cdot)/\{1 - \text{log}(\cdot)\}$}
    \begin{tabular}{ccc}
    \hline
         Specification 1 & Specification 2 & Abbreviation \\
         \hline
         %\hhline{|=|=|=|=|=|=|}
$\text{logit}\{ P(M_{ij} = 1) \} = \eta_{0} + \eta_{1}X_{ij} + \eta_{2}A_{i} + \eta_{3}Y_{ij}$ & $\text{logit}\{ P(M_{ij} = 1) \} = \eta_{0} + \eta_{1}X_{1ij} + \eta_{2}X_{2ij} + \eta_{3}X_{3ij} + \eta_{4}A_{i} + \eta_{5}Y_{ij}$ & $M \sim X + A + Y$ \\
\rowcolor{Gray}
$\text{logit}\{ P(M_{ij} = 1) \} = \eta_{0} + \eta_{1}X_{ij} + \eta_{2}A_{i} + \eta_{3}Y_{ij} + \eta_{4}A_{i}Y_{ij}$ & \begin{tabular}{@{}c@{}} $\text{logit}\{ P(M_{ij} = 1) \} = \eta_{0} + \eta_{1}X_{1ij} + \eta_{2}X_{2ij} + \eta_{3}X_{3ij} + \eta_{4}A_{i} + \eta_{5}Y_{ij} $ \\ $+ \eta_{6}A_{i}Y_{ij}$ \end{tabular} & $M \sim X + A*Y$ \\
$\text{logit}\{ P(M_{ij} = 1) \} = \eta_{0} + \eta_{1}X_{ij} + \eta_{2}A_{i} + \eta_{3}Y_{ij} + \eta_{4}X_{ij}A_{i}$ & \begin{tabular}{@{}c@{}}$\text{logit}\{ P(M_{ij} = 1) \} = \eta_{0} + \eta_{1}X_{1ij} + \eta_{2}X_{2ij} + \eta_{3}X_{3ij} + \eta_{4}A_{i} + \eta_{5}Y_{ij}$ \\
$+ \eta_{6}X_{1ij}A_{i} + \eta_{7}X_{2ij}A_{i} + \eta_{8}X_{3ij}A_{i}$ \end{tabular} & $M \sim X*A + Y$ \\
\rowcolor{Gray}
\begin{tabular}{@{}c@{}}$\text{logit}\{ P(M_{ij} = 1) \} = \eta_{0} + \eta_{1}X_{ij} + \eta_{2}A_{i} + \eta_{3}Y_{ij}$ \\ $+ \eta_{4}X_{ij}A_{i} + \eta_{5}A_{i}Y_{ij}$ \end{tabular} & \begin{tabular}{@{}c@{}}$\text{logit}\{ P(M_{ij} = 1) \} = \eta_{0} + \eta_{1}X_{1ij} + \eta_{2}X_{2ij} + \eta_{3}X_{3ij} + \eta_{4}A_{i} + \eta_{5}Y_{ij}$ \\
$+ \eta_{6}X_{1ij}A_{i} + \eta_{7}X_{2ij}A_{i} + \eta_{8}X_{3ij}A_{i} + \eta_{9}A_{i}Y_{ij}$ \end{tabular} & $M \sim X*A + Y*A$ \\
\begin{tabular}{@{}c@{}}$\text{logit}\{ P(M_{ij} = 1) \} = \eta_{0} + \eta_{1}X_{ij} + \eta_{2}A_{i} + \eta_{3}Y_{ij}$ \\ $+ \eta_{4}X_{ij}A_{i} + \eta_{5}A_{i}Y_{ij} + \eta_{6}X_{ij}Y_{ij} + \eta_{7}X_{ij}A_{i}Y_{ij}$ \end{tabular} & \begin{tabular}{@{}c@{}}$\text{logit}\{ P(M_{ij} = 1) \} = \eta_{0} + \eta_{1}X_{1ij} + \eta_{2}X_{2ij} + \eta_{3}X_{3ij} + \eta_{4}A_{i} + \eta_{5}Y_{ij}$ \\
$+ \eta_{6}X_{1ij}A_{i} + \eta_{7}X_{2ij}A_{i} + \eta_{8}X_{3ij}A_{i} + \eta_{9}A_{i}Y_{ij} + \eta_{10}X_{1ij}Y_{ij} + \eta_{11}X_{2ij}Y_{ij}$ \\ $+ \eta_{12}X_{3ij}Y_{ij} + \eta_{13}X_{1ij}A_{i}Y_{ij} + \eta_{14}X_{2ij}A_{i}Y_{ij} + \eta_{15}X_{3ij}A_{i}Y_{ij}$ \end{tabular} & $M \sim X*A*Y$ \\
\hline
    \end{tabular}
    \label{tab:one}
    \end{adjustwidth}
\end{sidewaystable*}
%\end{landscape}

Since both scenarios used binary effect modifiers, all imputation procedures were specified under the umbrella of logistic regression procedures, i.e., SI and MI used logistic regression, while the MMI and B-MMI methods used logistic mixed-effects models with a random intercept at the cluster level (B-MMI using estimates from such models as initial values). For the B-MMI method, each MCMC chain included 1000 burn-in iterations and used thinning to draw one set of posterior samples every 100 iterations after burn-in until $D=15$ complete data sets were collected, for a total of 2500 iterations. For all multiple imputation procedures, we combined the outcome model estimates based on each complete data set using Rubin's rule as described earlier.

\subsection{Performance Measures}

Four performance measures were considered for each simulation scenario and estimand. First, the bias for each method was calculated as the mean difference of estimates across simulation iterations and the corresponding true estimand value described in the Estimands section. As a second measure, the coverage was calculated as the proportion of simulations for which the estimated confidence interval contained the corresponding true estimand value. Third, the power was calculated as the proportion of simulations which would reject the null hypothesis (that the corresponding estimand is equal to 0). This is equivalent to Type \rom{1} error when simulating under the null. Finally, the mean squared error (MSE) was calculated as the average squared difference of estimates from each simulation and the corresponding true estimand value. The number of simulations run for each data-generating mechanism was chosen to ensure a reasonably low Monte Carlo standard error for the coverage estimates. In particular, supposing that the true coverage for a method is $95\%$, then approximately 1900 simulations for each data-generating mechanism are required to achieve a Monte Carlo standard error of $0.5\%$. This was then rounded up to 2000 simulation iterations. Thus, for a method that truly has 95\% coverage, we would expect to find estimated coverage between 94\% and 96\% across the vast majority of data-generating mechanisms we considered.

\subsection{Results}

The results for the HTE estimand when setting $\beta_{3} = \gamma_{3} = 0$ in the first scenario are presented in Figure 2. Figure 2 shows that SI and MI can dramatically reduce bias compared to CCA when using the three-way interaction imputation model. However, some bias remains. The MMI and B-MMI approaches reduce bias further and have near-zero bias under the three-way interaction imputation model, but exhibit similar bias under the other imputation models. As expected, coverage was far below the nominal level for the SI method, but this recovered greatly after performing MI. B-MMI under the three-way interaction imputation model was the only method to achieve approximately nominal coverage and maintain the specified $\alpha = 0.05$ Type \rom{1} error rate, but could sometimes lead to overcoverage or undercoverage under misspecified imputation models. MI, MMI, and B-MMI all provided comparable MSE. Results for the ATE estimand under this data-generating mechanism are presented in Web Figure 1 of the Supplemental Material. While all methods showed a small amount of bias and undercoverage for this estimand, MI, MMI and B-MMI were the closest to achieving zero bias and nominal coverage, with better performance than SI or CCA.

\begin{figure*}[htbp]
\begin{center}
\includegraphics[]{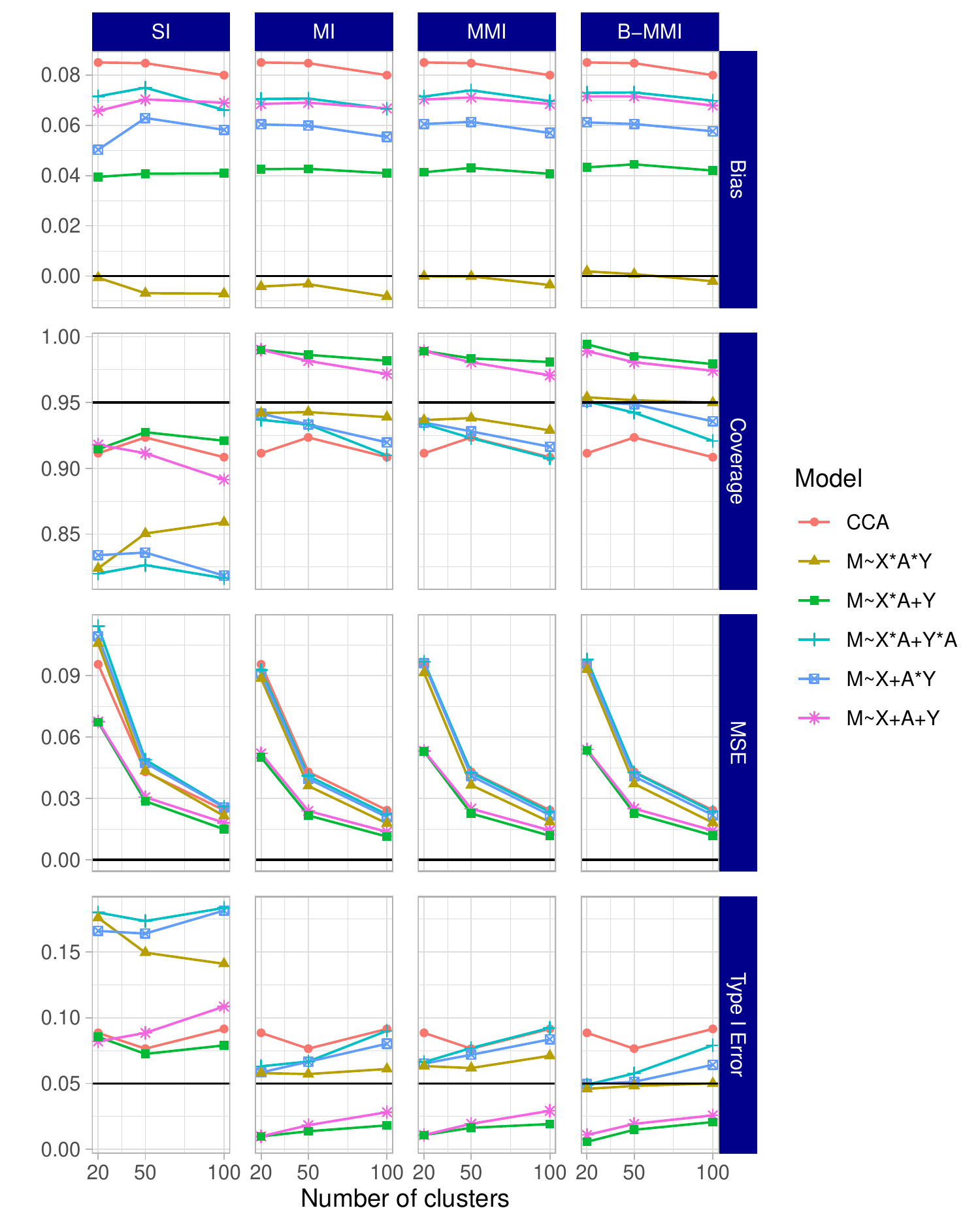}
\caption{Simulation results for the HTE estimand (interaction effect estimand $\gamma_{3})$ in the first simulation scenario when $\gamma_{3} = 0$. Rows from top to bottom show the bias, coverage, MSE, and Type I Error metrics. Columns from left to right show the performance of SI, MI, MMI, and B-MMI methods. CCA is displayed for comparison on each panel.}
\end{center}
\end{figure*}

Results for the HTE estimand and ATE estimand when setting $\beta_{3} = \gamma_{3} = -\{1 + \exp(-0.5)\}$ in Scenario 1 are presented in Web Figures 2 and 3, respectively. \textcolor{blue}{For the HTE estimand, B-MMI slightly outperformed MMI across all imputation model specifications, followed by MI and SI. However, the differences between B-MMI and MMI were minimal.} Notably, each of the imputation approaches can have substantially more bias than CCA if a highly incompatible imputation model is used, such as the imputation model with main effects only. Each imputation approach can also exhibit much worse coverage than CCA when the imputation model is misspecified, with coverage deteriorating further as the number of clusters increases; this is because the estimated confidence intervals narrow around biased point estimates. For the ATE estimand (Web Figure 3), performance was similar for MI, MMI, and B-MMI, with each achieving lower bias and closer to nominal coverage than SI or CCA, although a small amount of bias remained.

The results of the second simulation scenario were largely in agreement with the first scenario. The results for the HTE estimand $\gamma_{3}$ when setting $\beta_{3} = \gamma_{3} = 0$ in the second scenario are presented in Figure 3. All imputation methods had low bias under the three-way interaction imputation model, but performed poorly under highly misspecified imputation models. However in this scenario, B-MMI had noticeably higher coverage than MMI under all imputation models, and was the only method to achieve nominal coverage. Likewise, B-MMI under the three-way interaction model was the only method to achieve the nominal Type \rom{1} error rate. The results for the ATE estimand for this data-generating mechanism are presented in Web Figure 4. For this estimand, MI, MMI, and B-MMI all yielded good and comparable performance.

Finally, the results for the HTE and the ATE estimands when setting $\beta_{3} = \gamma_{3} = -\{1 + \exp(-0.5)\}$ in Scenario 2 are presented in Web Figures 5 and 6. For the HTE estimand $\gamma_{3}$, MMI and B-MMI exhibited lower bias and higher coverage than SI or MI across all specified imputation models. But B-MMI under the three-way imputation model was the only method to achieve nominal coverage. As in Scenario 1, all imputation methods could perform substantially worse than CCA under highly misspecified imputation models. For the ATE estimand, performance was similar for MI, MMI, and B-MMI, with low but non-zero bias and slight undercoverage. Overall, B-MMI consistently had the strongest performance in the simulation scenarios considered, followed by MMI and then MI. \textcolor{blue}{As noted in theoretical examinations in related work,\cite{meng1994,murray2018} Bayesian methods may perform better than their Frequentist counterparts in scenarios where imputation and outcome model specification are incompatible because the Bayesian approaches more appropriately handle uncertainty in parameter estimation. This feature likely explains the improvements we found when comparing B-MMI with MMI.}

\begin{figure*}[htbp]
\begin{center}
\includegraphics[]{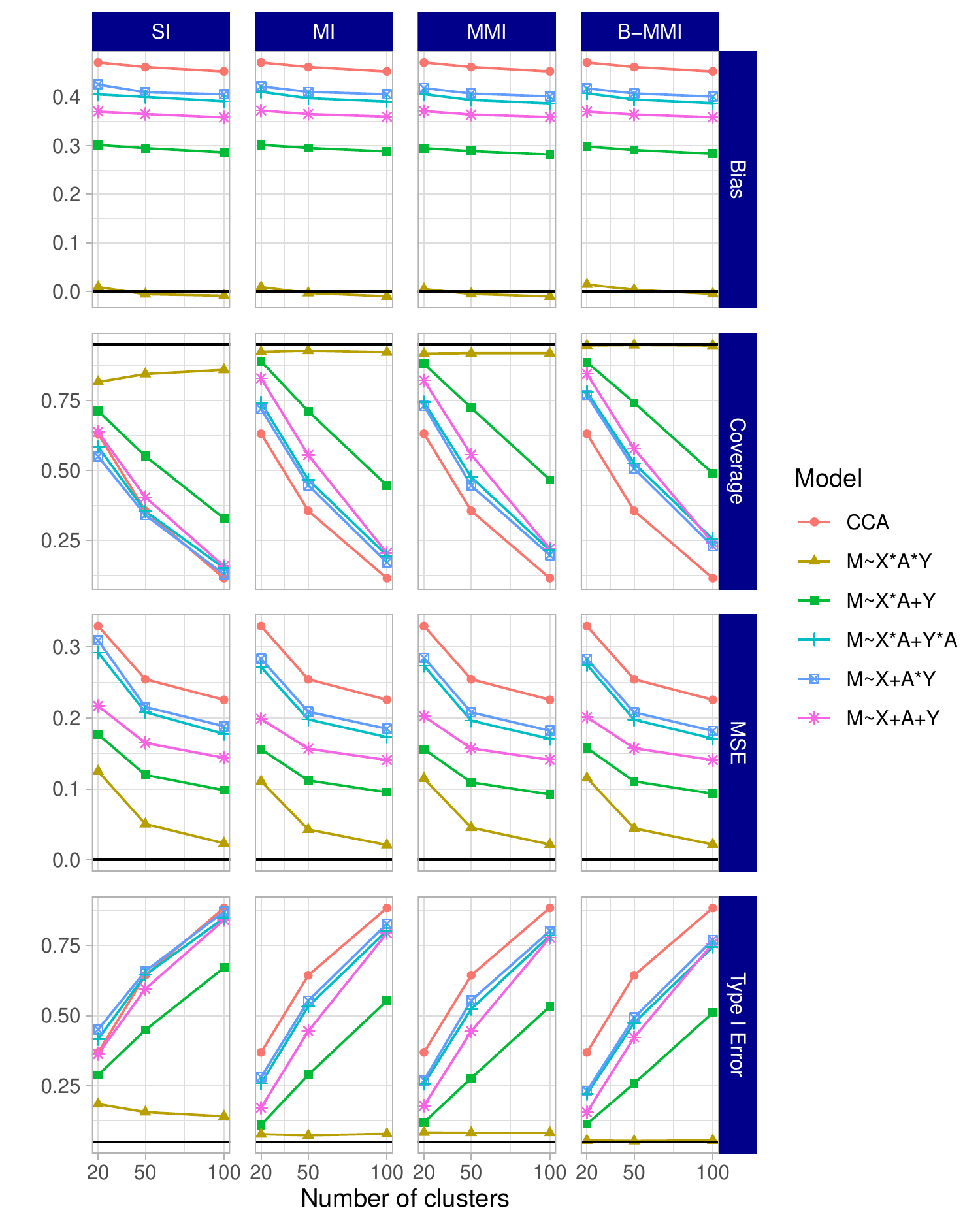}
\caption{Simulation results for the HTE estimand (interaction effect estimand $\gamma_{3}$) in the second simulation scenario when $\gamma_{3} = 0$. Rows from top to bottom show the bias, coverage, MSE, and Type I error metrics. Columns from left to right show the performance of SI, MI, MMI, and B-MMI methods. CCA is displayed for comparison on each panel.}
\end{center}
\end{figure*}

\section{Application to the Work, Family, and Health Study (WFHS)}

In this section, we illustrate each of the methods described earlier by analyzing %publically available 
data from the Work, Family, and Health Study (WFHS). WFHS consisted of two CRTs conducted at two employers; here, we focus only on the experiment conducted at one employer, an extended-care company.\cite{wfhs2018} The trial included 30 work sites of 30-89 employees each. The work sites were randomly assigned to a comprehensive work-family intervention or usual practice conditions. \textcolor{blue}{The intervention consisted of two main components, one to increase managers' support for their employees' family and personal lives, and one to improve employees' control over their schedule. As managers already have more control over their schedule and would be both giving and receiving the first component of the intervention, it's possible that the intervention effect on various outcomes varies by employee type (manager vs. non-manager).}

We specifically focus on this effect modification problem \textcolor{blue}{even though the employee type variable is completely observed,} such that we can simulate missing data scenarios while comparing to results from the known full data. In particular, we will study employee type as a potential modifier of the effect of the intervention on the outcome of time adequacy (TA), \textcolor{blue}{measured for each individual employee.} TA was a composite variable averaging the \textcolor{blue}{individual} survey responses to several questions of the form ``\emph{To what extent is there enough time to...}'' followed by a familial activity or responsibility. TA ranged from 1 to 5 with 5 signifying always having adequate time for family. \textcolor{blue}{Thus, TA acts as a proxy variable for the desirable but unmeasurable outcome of ``good work-life balance''.} Our outcome specification was the difference between TA at 12 months post-intervention and TA at baseline. This research question largely falls under Aim 4 of the selected sub-study of WFHS, to ``\emph{test whether employee, mid-level manager, and work- group characteristics moderate the effect of the intervention on work-family conflict and health outcomes.}''

Although the employee type variable is fully observed, we considered three hypothetical missing data scenarios for this effect modifier. In the first scenario (Figure 4(a)), about 20\% of the values of employee type were set to be missing completely at random. In the second scenario (Figure 4(b)), about 20\% of the values of employee type were set to be missing at random as a simple function of the outcome and two additional covariates: self-reported job autonomy and a self-reported assessment of control of schedule, each of which was reported on a Likert scale of 1 to 5, with 5 indicating more control/autonomy. In particular,
\begin{align*}
    \text{logit}\{ P(\text{Employee Type is not missing}) \} = 2 &+ 0.5TA - 0.6(\text{Control of Schedule} \geq 4) \\
    &- 0.3(\text{Job Autonomy} \geq 4)
\end{align*}
The third scenario is a similar MAR setup (Figure 4(b)) to the second scenario, but missingness was simulated as a more complicated function of outcome and covariates, including a random effect to induce clustering of missingness within work sites.
\begin{align*}
    \text{logit}\{ P(\text{Employee Type is not missing}) \} = \zeta &+ 2 + 0.5TA - 0.6(\text{Control of Schedule} \geq 4) \\
    &- 0.3(\text{Job Autonomy} \geq 4) \\
    &+ 0.05TA(\text{Control of Schedule} \geq 4) \\
    &- 0.15TA(\text{Job Autonomy} \geq 4) \\
    &+ 0.1TA(\text{Control of Schedule} \geq 4)(\text{Job Autonomy} \geq 4)
\end{align*}
where $\zeta$ is a random intercept at the work site level generated as a Normal variable with mean 0 and variance such that ICC on the latent response scale was equal to 0.1.

\begin{figure*}[htb]
\begin{center}
% \fbox{\includegraphics[width=5.5in, trim={3.5cm 8.5cm 3cm 8.5cm}]{DAGS_for_HTE_paper_3.pdf}}
\fbox{\includegraphics[scale=0.65, trim={3.5cm 8.5cm 3cm 8.5cm}]{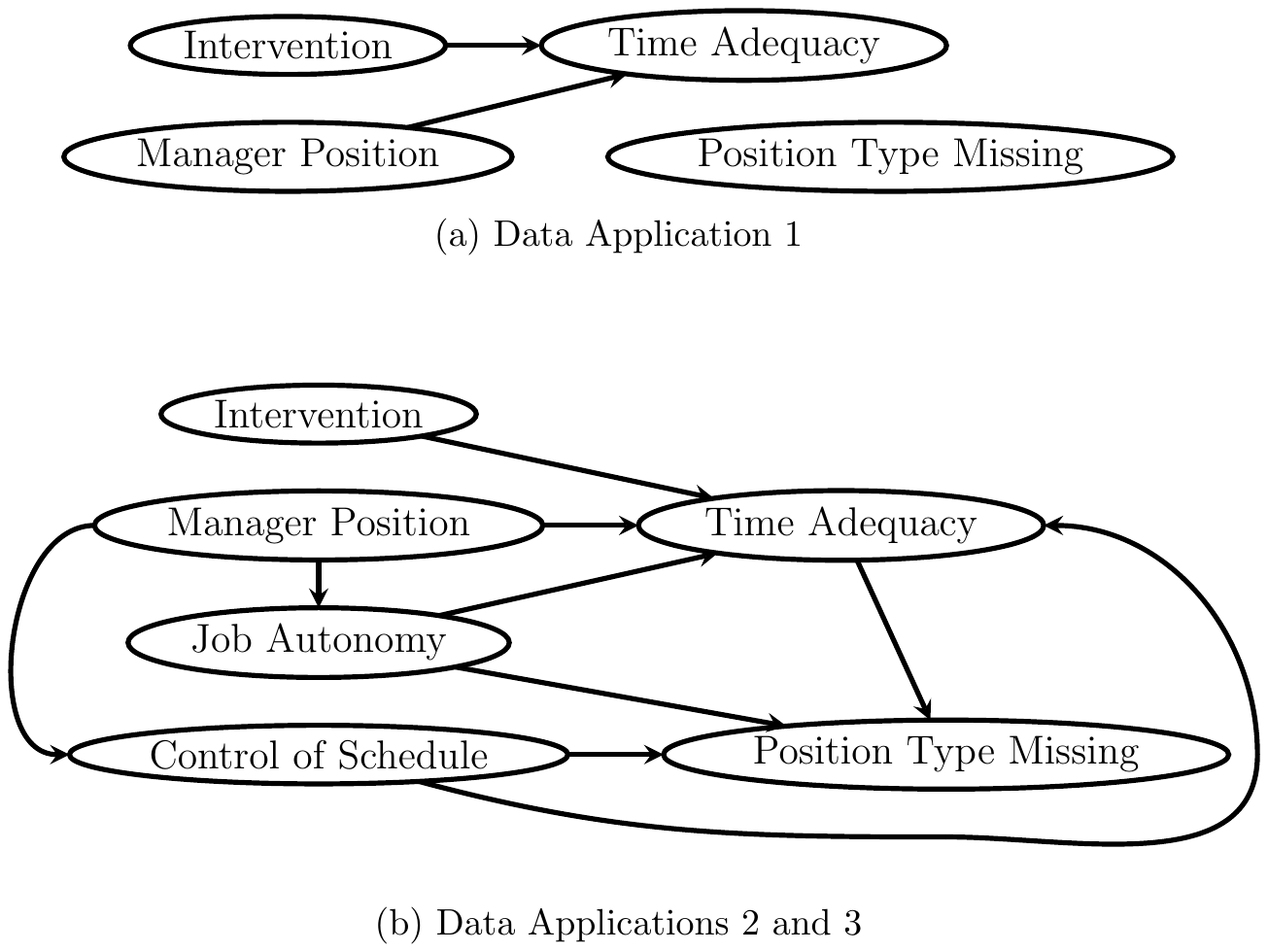}}
\caption{Directed acyclic graphs (DAGs) corresponding to the Work, Family, and Health Study data application scenarios that were considered.}
\end{center}
\end{figure*}

In each scenario, the 5 methods are compared by calculating the point estimates and confidence intervals for the two estimands described earlier - the coefficient for interaction between intervention and employee type (the heterogeneous treatment effect) and the average treatment effect. \textcolor{blue}{As we mentioned earlier, we assume the absence of informative cluster size and a correct specification of the GEE marginal mean model for data analysis; therefore the average treatment effect and the interaction effect can be mapped to the corresponding regression coefficients and interpreted without ambiguity.
%Although both a participant-average and cluster-average effect could be of interest in this trial, we focus on the participant level estimand to remain consistent with the rest of the article.
} For the MAR scenarios, each imputation method specified an imputation model that included the outcome, the two baseline covariates (dichotomized as in the missing data generation), the treatment, and all possible two-way interactions between them. Models with higher-order interaction terms did not always converge due to the limited sample size and nontrivial number of parameters, and were thus excluded. \textcolor{blue}{The missing data simulation procedure and corresponding estimation under each method was repeated for a comparison across 500 replications.}

\begin{figure*}[htbp]
\begin{center}
\includegraphics[width=5.5in]{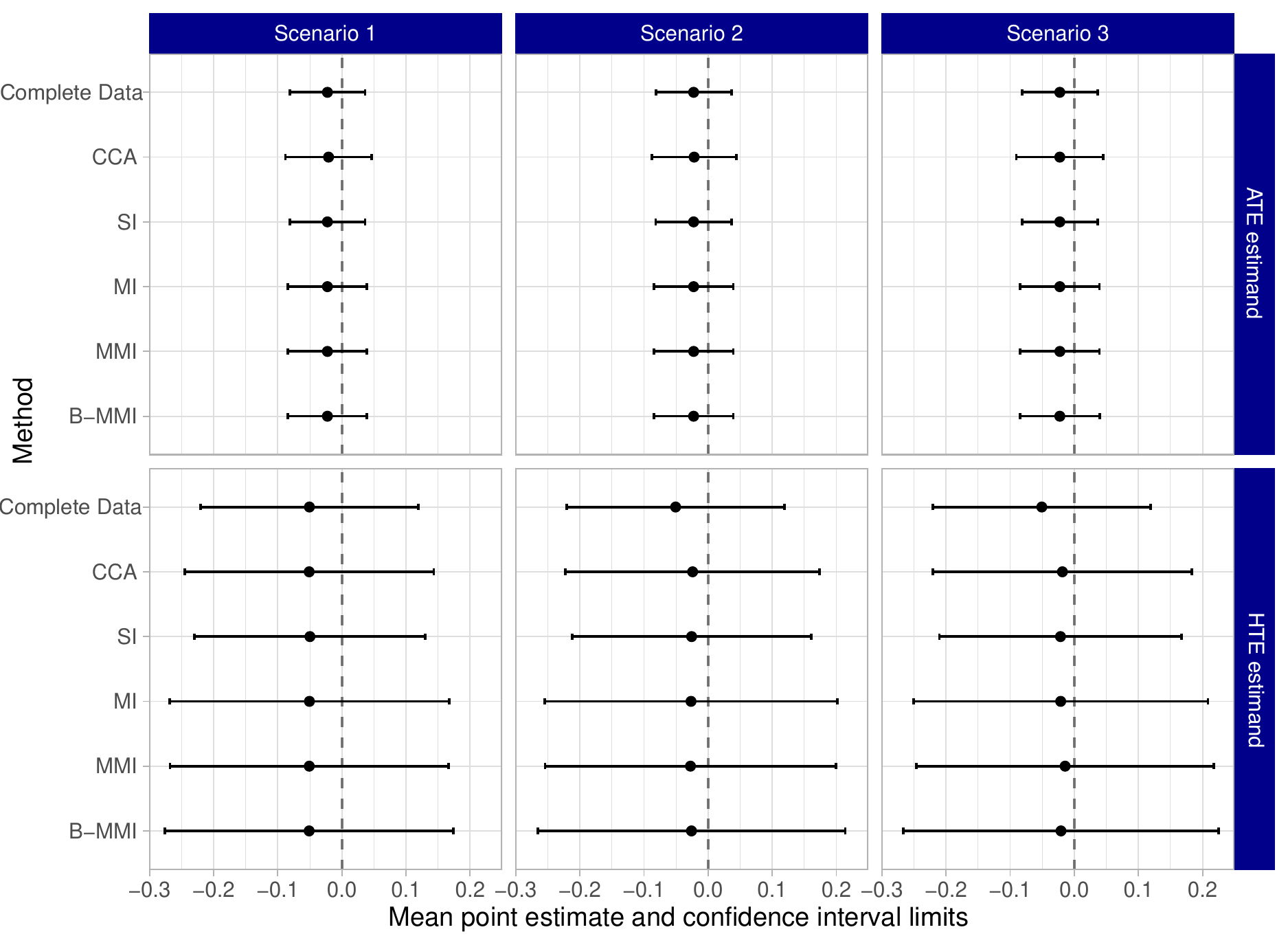}
\caption{Results for the Work, Family, and Health Study data application. The top row shows the average point estimate and average confidence interval limits for each method across 500 replications of the simulation procedure for the ATE estimand. The bottom row shows the parallel results for the HTE estimand.}
\end{center}
\end{figure*}

\textcolor{blue}{Key results are presented in Figure 5, which shows the average point estimate and average 95\% confidence (upper and lower) limits across the 500 replications for each method. For these results, 13 iterations were removed in Scenario 2 for MMI due to the non-convergence of the imputation model; these were resolved for B-MMI by using alternative initial values informed by the SI/MI imputation model. For the ATE estimand (top row of the figure), all methods seem to perform similarly but with slight variation in average confidence interval width.} For the HTE estimand \textcolor{blue}{(bottom row of the figure)}, CCA yielded similar results to each of the imputation methods under all scenarios, and produced narrower confidence intervals on average than MI, MMI, or B-MMI. This is in slight contrast to the results from the simulation studies, and may be due to a lack of explanatory power of the assumed imputation model with the available auxiliary data. 
%ability to predict the modifier well with available auxiliary data. 
\textcolor{blue}{Alternatively, this phenomenon can arise as a result of increased uncertainty in the imputation procedure. That is, whereas CCA uses less data, 
%which may increase confidence interval width compared to methods that use more data, another possible explanation for this phenomenon is that CCA fails to account for uncertainty in the missingness mechanism, which reduces CI width compared to methods which make a MAR assumption under a similar effective sample size. 
the multiple imputation methods can introduce nontrivial uncertainty around the imputed values and therefore increase the width of the confidence interval. Finally, although SI had similar confidence limits to the complete data on average, SI was also the only method for which the averaged confidence intervals did not fully cover the range of the complete data confidence interval in Scenarios 2 and 3. MI, MMI, and B-MMI performed similarly, but had noticeably wider average confidence intervals than CCA or SI. Although there is no ground truth to compare to, this may indicate that the set of multiple imputation procedures are more faithfully accounting for the uncertainty due to missing effect modifier data.}

\textcolor{blue}{Web Tables 1 and 2 report the percentage of iterations where each methods' confidence intervals were narrower than the corresponding complete data confidence interval, for the ATE and the HTE estimands respectively. %Such narrowness in output may indicate overconfidence of the method. 
Notably, MI, MMI, and B-MMI almost never produced a narrower confidence interval compared to the complete data confidence interval, except for at most two iterations in a given scenario. In contrast, CCA yielded a narrower confidence interval compared to the complete data counterpart between 9.0\% and 16.2\% of the time, depending on the scenario. Furthermore, SI reported a narrower confidence interval than the complete data counterpart between 22.2\% and 48.6\% of iterations across all scenarios. This adds evidence to our Simulation Study section results that SI often yields overly narrow confidence intervals. Web Tables 3 and 4 report a related metric, the percentage of iterations where each methods' 95\% confidence interval completely covered the complete data 95\% confidence interval. 
%While this may not be straightforward to interpret, 
A very low value of this metric could indicate that an approach is frequently excluding plausible parameter values. For the ATE estimand, MI, MMI, and B-MMI almost always covered the complete data interval, while CCA and SI tended to cover the complete data interval only around a third of the time. For the HTE estimand in Scenario 2, CCA, SI, MI, MMI, and B-MMI reported covering the complete data confidence interval 34.0\%, 19.4\%, 60.6\%, 63.2\%, and 69.4\% of the time, respectively. Results for other scenarios were similar, and collectively indicate that SI likely did not appropriately account for uncertainty, while B-MMI appropriately accounts for more uncertainty on average than other approaches.}

\textcolor{blue}{Finally, to further investigate the variability of the point estimates across replications, we provide box plots of the point estimates in Web Figure 7. The complete data point estimate and 95\% confidence interval are presented on the left of each panel for reference. The top row displays results for the ATE estimand, where CCA exhibited much higher variability than the other approaches. SI exhibited slightly higher variability than the other imputation approaches. All methods' median point estimates were close to the complete data estimate. The bottom row of the figure displays results for the HTE estimand, and suggests that no method had a median point estimate equal to the complete data point estimate in Scenario 2 or 3. While all methods tended to report point estimates within the complete data confidence interval, each had point estimates outside this range in either Scenario 2 or 3. As expected, SI has highly variable point estimates, with several point estimates outside of the complete data confidence interval. MMI reports a bias-variance tradeoff in Scenario 3, with its median point estimate being the furthest from the complete data point estimate, but also having less variable estimates overall.}

\textcolor{blue}{We acknowledge that there is no known truth in this single data application, and the complete data results are themselves subject to variability. Thus, whether or not confidence intervals ``match'' those for the complete data should only be interpreted with caution. However, when considering all metrics together, this application provides strong evidence that CCA and SI are likely overly confident, and that MMI and B-MMI may be most faithful in capturing uncertainty. For the complete data, both the ATE and the HTE estimates were close to null with relatively wide confidence intervals, indicating no average treatment effect and no effect heterogeneity. Although it is difficult to define the scale of meaningful treatment effects when the outcome is an average of Likert variables, the confidence intervals exclude treatment effects of an absolute value larger than 0.1, which would correspond to a very minor difference between groups.}

\section{Discussion}

In this article, we compared several methods to address missing effect modifier data for assessing HTE (and for assessing ATE as a direct product of the analysis model) in CRTs using simulation studies and then used a completed CRT where we imposed selected missing data patterns to compare their performance using real data. The key findings of our work are as follows. First, MMI and B-MMI had the lowest bias and highest coverage across the settings we investigated, with B-MMI being the only method to achieve nominal coverage rates across several scenarios considered when the imputation models were approximately correct. However, when imputation models were strongly misspecified, imputation approaches performed poorly, and could even be worse than CCA in several scenarios. Second, using an imputation model with only main effects for the outcome, treatment, and covariates resulted in especially poor performance when targeting a non-null HTE estimand. This could have important implications in practice as imputation models with only main effects are the default specification in many software packages, such as the popular \texttt{mice} R package.\cite{van2011} \textcolor{blue}{Third, including more interaction terms almost always resulted in better performance in the simulations, so in practice, it may be safer to ``overspecify'' imputation models when enough data is available to justify doing so.} Future work should consider whether such overspecifications maintain good performance when the smallest necessary imputation model is more parsimonious. Finally, implementing the logistic mixed-effects model Gibbs Sampler for B-MMI is, to our knowledge, a new contribution; previous procedures in the multilevel data context have instead leveraged a probit link function.

The simulation study focused on data-generating mechanisms that induced correlations within clusters, as are commonly found in CRTs. Accounting for these correlations when performing imputation was critical, and MMI and B-MMI generally outperformed the other comparator methods throughout the simulation studies. In CRTs, much attention has been given to accounting for intracluster correlation of outcome variables, but similar consideration must also be given for the intracluster correlation of covariates, especially for the purpose of studying confirmatory HTE. The recommendation to account for correlations in effect modifiers in CRTs has been previously emphasized for designing CRTs,\cite{yang2020,li2022designing,maleyeffsample,ryan2023maximin,tong2023sample,tong2022,tong2023accounting} and here we have reinforced that same recommendation when imputing missing effect modifier data in CRTs. This recommendation is more related to correct imputation model specification than to model compatibility. In this manuscript, the terms ``model misspecification'' and ``lack of model compatibility'' were used somewhat interchangeably, but there are subtle differences between them. While specification of interaction terms in imputation models is important for compatibility, adjusting for correlations in a missing effect modifier may be primarily needed to reflect the extra variability in the effect modifier at the cluster level and thus to ensure valid uncertainty statements when analyzing the imputed complete data. \textcolor{blue}{While we primarily addressed missing effect modifier data at the individual level due to its strong relevance to mainstream practice in subgroup analysis of CRTs,\cite{wang2023sample} future work should also consider the impact of missing modifiers which are measured at the cluster level rather than the individual level.} %Because modifiers are already conditioned on in the outcome model, the random effects in modifier imputation models may not be necessary for compatibility purposes.

A summary of findings and recommendations for practitioners reflecting our observations is provided in Table 2. \textcolor{blue}{As always, assessing the plausibility of MCAR and MAR assumptions should be the first step in addressing missingness. Then one should consider the assumptions of different methods and model specifications, noting that while each imputation method assumes MAR, they are distinguished by whether they account for partial or more complete uncertainty in the imputation process. These distinctions were informed by the simulations and real data analysis, where some methods were more robust to specifying imputation models which were incompatible with the substantive model.} Overall, B-MMI has been shown to be a promising approach for handling missing effect modifier data in CRTs, and generally had the strongest performance among the methods that we considered. 
%Practitioners with experience coding Gibbs samplers should be able to implement this method in practice. 
To facilitate implementation, all simulation and data application code are available at \url{https://github.com/harhay-lab/CRT-miss-mod}. For practitioners that do not perform, or prefer, analysis under the Bayesian paradigm, approximate (Frequentist) MMI also exhibited acceptable performance for missing modifier data in CRTs in several scenarios. However, there may be bias or non-nominal coverage when using MMI or B-MMI in similar ways that we have in this article, e.g., when the imputation models are not truly compatible with the outcome model. This was seen for the ATE estimand, where no method achieved nominal coverage or preserved the Type \rom{1} error rate in Scenario 1 of the simulation study. Although using three-way interactions in the imputation models was often sufficient for good performance, there does not exist a joint model for which the implied conditional outcome model is a linear model with random intercept and the implied conditional modifier model is a logistic model with random intercept. This highlights the importance of and potential challenge in the imputation model specification in CRTs when the interest lies in assessing HTE (and ATE after effect modifier adjustment) with a single binary effect modifier. These recommendations will not necessarily be generalizable to other settings such as scenarios with continuous effect modifiers or binary outcomes, and future work in necessary to expand our studies to those settings.

\begin{table*}[]
    %\textwidth
    \footnotesize
    \caption{Summary of findings from the simulation and data application comparisons and recommendations for practitioners}
    \begin{tabular}{l}
    \hline
         $\cdot$ Model compatibility between the outcome and imputation models is key for unbiased estimation and inference. \\[0.075cm]
         \rowcolor{Gray}
         $\cdot$ As assessing HTE often entails interaction terms in outcome models, corresponding interaction terms
         should be used in \\
         \rowcolor{Gray}
         imputation models. \\[0.075cm]
         $\cdot$ Cautiously over-specifying imputation models may be a good strategy when enough data is available for the models to \\
         converge. \\[0.075cm]
         \rowcolor{Gray}
         $\cdot$ As with missing outcome data, correlation within clusters should be accounted for when imputing missing modifier data. \\[0.075cm]
         $\cdot$ MMI and B-MMI are promising methods in this setting and recommended over CCA, SI, or MI, but may still have bias or \\ undercoverage when using non-compatible models. \\[0.075cm]
         \rowcolor{Gray}
         $\cdot$ B-MMI is recommended when targeting an HTE or interaction effect estimand. \\[0.075cm]
         $\cdot$ MMI and B-MMI performed similarly and are recommended when targeting an ATE estimand. \\[0.075cm]
         %\rowcolor{Gray}
         %$\cdot$ Sensitivity analyses including CCA and one or more imputation methods should be used in practice \\
\hline
    \end{tabular}
\end{table*}

%This highlights that further research that is beyond the scope of this initial contribution is necessary. In particular, 
There are several paths forward that have been illuminated by these comparison studies. First, ``substantive-model compatible'' approaches defined for individually-randomized trials or observational studies\cite{goldstein2014,bartlett2015,kim2015,enders2020,ludtke2020} could be compared and may be valuable tools in the CRT setting. These often entail deriving the correct imputation model once an outcome model is specified, and (as the correct model will likely have a non-standard form) sampling from this derived model using MCMC. Joint modeling methods for individually-randomized trials and observational studies\cite{zhang2017,kim2018,erler2019} may be similarly useful. Some of these are likely adaptable to the CRT setting, while others may require extension as they are either not designed for a multilevel setting, do not have off-the-shelf software, and/or are dependent on specific outcome model forms. Second, the use of non-parametric imputation approaches that are flexible enough to be compatible with a large range of outcome models also merit deeper examination. One procedure that may be especially useful for this is imputation via Bayesian Additive Regression Trees, or BART.\cite{chipman2010,xu2016,chen2022bayesian} Not only would such methodology be useful for handling missing modifiers in CRTs, but it is also under-explored in individually-randomized trial and observational study settings where studying HTE introduces similar model compatibility issues. If such an approach can perform well without requiring the expertise to both derive an implied imputation model and code an appropriate sampling procedure, it would be practical for assessing HTE with missing modifier data. We plan to pursue this methodological extension in a separate report.

Our studies have several limitations. First, although we considered a wide range of data-generating mechanisms, there are inevitably many possible data structures that were not included in the simulation studies. \textcolor{blue}{For example, all data generating mechanisms generated modifier data independently of the baseline covariates. While this still implies a complex imputation model, as described in the Simulation Study, it leaves out plausible scenarios where auxiliary covariates have a direct impact on the modifier. Previous work indicates that such correlations may not greatly impact simulation results,\cite{seaman2012} but this still may be an interesting area for future exploration.}
%Only particular linear outcome model forms were considered, and results may be different under different models or for discrete outcome variables. Furthermore, 
\textcolor{blue}{Furthermore, we restricted our comparisons to scenarios with one individual-level binary effect modifier, which are common in practice when assessing subgroup-specific treatment effects or HTE.\cite{wang2023sample} Future work should consider method performance when modifiers are categorical or continuous variables, when effect modifiers are measured at the cluster level, or when there are multiple effect modifiers that are subject to missingness. In addition, for an individual-level effect modifier, varying the ICC of the effect modifier may be useful in assessing the operating characteristics of the MMI and B-MMI methods.} 
%We varied number of clusters as a key variable of interest, but varying the ICCs or number of random effects would also be informative. 
To address data with multiple missing variables in CRTs, imputation via joint modeling and imputation via fully conditional specification should also be compared in future research; see Audigier et al.\cite{audigier2018} for such a comparison in the observational study setting. In general, the biases found should only amplify in scenarios with more than one effect modifier, such that our results are still informative and reflective of the recommended practice and caveats in that setting. Second, only MCAR and MAR scenarios were considered across the simulations and data illustration. Imputation-based approaches may not perform as well on data where modifiers are MNAR, and sensitivity methods that account for the multilevel feature of the CRT data may be useful to address MNAR. Third, in the simulation study, standard robust standard errors were used for all GEE outcome models. However, previous research has shown that bias-corrected standard error estimates are often needed when there are fewer than 30 clusters, especially for estimating a cluster-level effect parameter with missing outcomes.\cite{kauermann2001,mancl2001} In our simulation results with missing effect modifier data, the standard robust standard error estimator seemed adequate for as few as 20 clusters, but using bias-corrected standard errors may improve the slight undercoverage for the ATE estimand in a few settings. 
%Future work should investigate whether such bias corrections are helpful for data with a low number of clusters. 
%In the data application, a lower sample size (30 clusters with on average less than 40 participants each) led to non-convergence of more complex imputation models. Thus, the performance of imputation methods may have been hampered in the data illustration due to using overly simple models. 
Finally, while this article focused on imputation approaches, other methods for handling missing data in CRTs such as weighting-based methods\cite{turner2020} may be useful and require further developments to address missing effect modifier data in the CRT setting.\\

\noindent {\bf{Acknowledgements}}

\noindent Research in this article was partially supported by the Patient-Centered Outcomes Research Institute\textsuperscript{\textregistered} (PCORI\textsuperscript{\textregistered} Awards ME-2020C1-19220 to M.O.H. and ME-2020C3-21072 to F.L.). M.O.H. and F.L. are also funded by the United States National Institutes of Health (NIH), National Heart, Lung, and Blood Institute (grant number R01-HL168202). All statements in this report, including its findings and conclusions, are solely those of the authors and do not necessarily represent the views of the NIH or PCORI\textsuperscript{\textregistered} or its Board of Governors or Methodology Committee. We would also like to thank the investigators and participants of the Work, Family, and Health Study. The data from the Work, Family, and Health Study is publicly available at \url{https://doi.org/10.3886/ICPSR36158.v2}.

\bibliographystyle{SageV}
\bibliography{refs.bib}

\end{document}